\documentclass[12pt,a4paper]{article}
\usepackage{jheppub}

\usepackage{graphicx}
\usepackage{subcaption}
\usepackage{xspace}

\usepackage[utf8]{inputenc}

\usepackage[style=numeric-comp,sorting=none]{biblatex}
\abstract{Data processing frameworks are an essential part of HEP
experiments' software stacks. Frameworks provide a means by which code
developers can undertake the essential tasks of physics data processing,
accessing relevant inputs and storing their outputs, in a coherent way without
needing to know the details of other domains. Frameworks provide essential
core services for developers and help deliver a configurable working
application to the experiments' production systems.
Modern HEP processing frameworks are in the process of adapting to a new
computing landscape dominated by parallel processing and heterogeneity,
which pose many questions regarding enhanced functionality and scaling
that must be faced without compromising the maintainability of the code.
In this paper we identify a program of work that can help further clarify
the key concepts of frameworks for HEP and then spawn R\&D activities that
can focus the community's efforts in the most efficient manner to address
the challenges of the upcoming experimental program.
}

\begin{document}

\noindent
\begin{tabular*}{\linewidth}{lc@{\extracolsep{\fill}}r@{\extracolsep{0pt}}}
 & & HSF-CWP-2017-08 \\
 & & December 19, 2018 \\ 
 & & \\
\end{tabular*}
\vspace{2.0cm}

\title{HEP Software Foundation Community White Paper Working Group -- Data Processing Frameworks}

\author{HEP Software Foundation:}
\author[d]{Paolo Calafiura}
\author[a]{Marco Clemencic}
\author[b]{Hadrien Grasland}
\author[c]{Chris Green}
\author[a,e,1]{Benedikt Hegner}
\author[c]{Chris Jones}
\author[b]{Michel Jouvin}
\author[c]{Kyle Knoepfel}
\author[g]{Thomas Kuhr}
\author[c,1]{Jim Kowalkowski}
\author[d]{Charles Leggett}
\author[c]{Adam Lyon}
\author[e]{David Malon}
\author[c]{Marc Paterno}
\author[d]{Simon Patton}
\author[c,1]{Elizabeth Sexton-Kennedy}
\author[a]{Graeme A Stewart}
\author[d]{Vakho Tsulaia}



\affiliation[a]{CERN, Geneva, Switzerland}
\affiliation[b]{LAL, Université Paris-Sud and CNRS/IN2P3, Orsay, France}
\affiliation[c]{Fermi National Accelerator Laboratory, Batavia, Illinois, USA}
\affiliation[d]{Lawrence Berkeley National Laboratory, Berkeley, CA, USA}
\affiliation[e]{Brookhaven National Laboratory, Upton, NY, USA}
\affiliation[f]{Argonne National Laboratory, Lemont, IL, USA}
\affiliation[g]{Ludwig-Maximilians-Universität München, Munich, Germany}
\affiliation[1]{Paper Editor}

\maketitle

\newpage

\section{Introduction}
The aim of this document is to promote a common vision and roadmap for
data processing frameworks that will allow enhanced collaboration
across experiments and will meet future challenges given the variety
of stakeholders which will be defined below. We describe the problem
domain and introduce the concepts and relationships needed to
formulate common data processing framework solutions for the future.

The time periods of interest for this document are DUNE and HL-LHC,
which will deliver on the order of 50 PB of data events per year per
experiment. The results of the proposed R\&D ought to be used for
building the final software systems that will be utilized in
commissioning and operations of these experiments and the processing
of the data.

\section{Scope and Challenges}
\label{sec:scope-challenges}

Frameworks in HEP are used for the collaboration-wide data processing
tasks of triggering, reconstruction, and simulation, as well as other tasks that
subgroups of an experiment collaboration are responsible for, such as
detector alignment and calibration.
Providing common framework services and libraries that will meet with
compute and data needs for HL-LHC experiments and the Intensity Frontier
experiments is a large challenge given the multi-decade legacy in this
domain. At the same time the computing landscape is changing requiring
enhanced framework functionality to help users adapt to these changes.
We see a number of upcoming challenges that need to be addressed in the
coming decade:

\begin{enumerate}
\item
    Changes needed in the programming model that are necessary to
    handle the massive parallelism that will be present throughout all
    layers in the available computing facilities. This is necessary
    because of the ever-increasing availability of specialized compute
    resources, including GPGPUs, Tensor Processing Units (TPUs),
    tiered memory systems integrated with storage, and ultra
    high-speed network interconnects.
\item
    Challenges related to advanced detector technology, like finer
    granularity, high bandwidth continuous readout DAQ, and the need
    for a very tight feedback loop for conditions, which in turn
    requires a focus on low-latency solutions.
\item
    Tighter integration with the computing model: informing
    higher-level systems (workflow, data management, workload
    management, job management) based on fundamental changes in
    compute facilities.
\item
    Support structures that permit frameworks to be collaboratively
    developed and maintained across a large number of
    experiments. This includes excellent integration with the wider
    ecosystem encompasses development, deployment, and runtime
    components.
\item
    Provide flexibility and the interfaces required to efficiently
    integrate the various parallelization efforts on simulation,
    reconstruction and other fields.
\end{enumerate}

The challenge present in all of these elements is to ensure
productivity given the increased complexity and scale of the upcoming
experiments and computing facilities. This means decreasing program
development and debugging time, and increasing efficiencies in diverse
facilities use.  Frameworks have accomplished these tasks in the past,
the challenge is to continue this into the next generation experiments
and facilities.

\subsection{Stakeholders}
\label{sec:stakeholders}

\begin{figure}[hbt] 
\centering
\includegraphics[width=.75\textwidth]{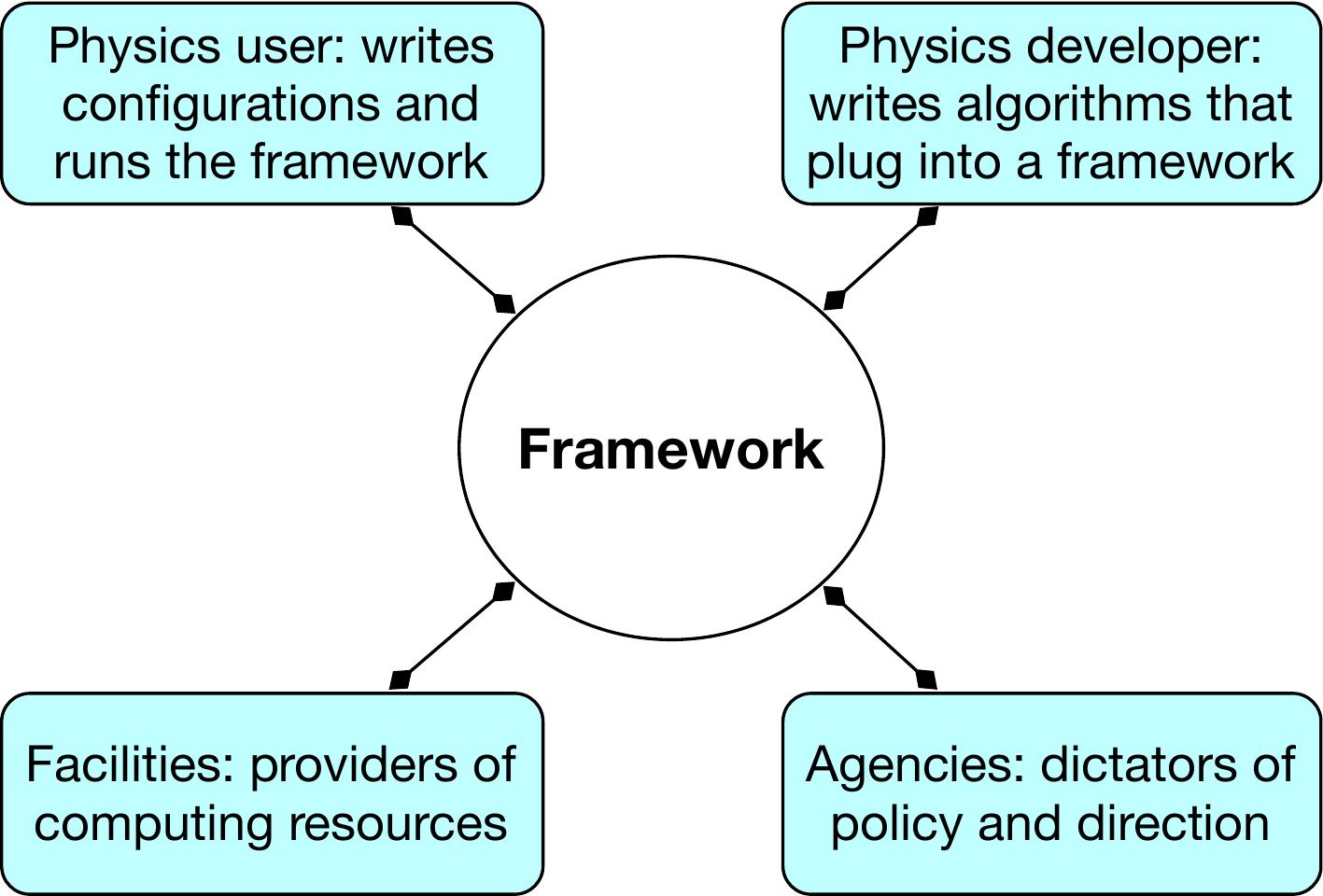}
\caption{Our stakeholders. \label{fig:stakeholders}}
\end{figure}

Understanding who influences the organization and behavior of the
framework is obviously important. This is typically accomplished by
collecting requirements and use cases. This paper does not pretend to
do this. We do, however, recognize the stakeholders that heavily
influence the requirements placed upon an event processing
framework. In this document we recognise that there are needs from
several kinds of stakeholders depicted in
Figure~\ref{fig:stakeholders} and listed below.  We attempt to address
the needs of all of these stakeholders.

\begin{enumerate}
\item
    Physics developers who write trigger, reconstruction, simulation,
    and analysis algorithms that plug into the framework. This
    category also includes providers of generalized infrastructures
    which are used by the physics developers to implement the above
    applications. Examples are the upgraded Geant simulation engine
    and online triggering infrastructures.These are important because
    interfacing to them may impose special requirements on the
    Framework.
\item
    Physics users who run framework programs for particular
    applications to produce collaboration wide datasets.
\item
    Funding agencies/laboratories, who mandate security requirements,
    or dictate the need for shared software infrastructure projects
\item
    Facilities who buy, operate, and build new systems that existing
    software infrastructures must be adapted to or rethought for
\end{enumerate}

\section{Current Practice}
\label{sec:current-practice}

Although most frameworks used in HEP share common concepts, there are,
for mainly historical reasons, a number of different implementations;
some of these are shared between experiments. A complete description
of framework use cases written for the Gaudi collaboration is
described in \cite{Maley:684466} and these are sufficiently general to apply
to all HEP experiments. The Gaudi framework \cite{Barrand:2001ny,1742-6596-219-4-042006} was originally
developed by LHCb, but is also used by ATLAS and various non-LHC
experiments. CMS uses its own CMSSW framework~\cite{Bayatian:2006nff}
which was forked to provide the art framework for the Fermilab
Intensity Frontier experiments~\cite{Green:2012gv}.  Belle II uses
basf2~\cite{1742-6596-331-3-032024}. The linear collider community
developed and uses Marlin~\cite{Gaede:2006pj}. The FAIR experiments
use FairROOT, closely related to ALICE's AliROOT. The FAIR experiments
and ALICE are now developing a new framework, which is called
ALFA~\cite{ALFA:ALICE-FAIR:2015}. At the time of writing, most major frameworks support
basic parallelisation, both within and across events, based on a
task-based model~\cite{Jones:2015soc,Clemencic:2015paa}. ALFA already
includes additional multi-node setups and communication.

The frameworks provide the necessary functionality like I/O,
scheduling, configuration, logging, etc.\ to support the execution of
these processing components. The aforementioned components provide
functionalities like pattern finding in a certain sub-detector or the
high-level identification of a given particle type. This layout allows
independent development and a high flexibility in the usage of physics
algorithms within an experiment collaboration.

The above defines frameworks in terms of its use. A more formal
definition is the framework holds the protocols, tools and concepts
for defining, developing, and deploying physics algorithms, along with
all the ancillary data and tools for providing services to the
algorithms.  This includes algorithm scheduling components, the event
data model, handling of input and output from physics applications
that utilize the framework, interfaces for non-event data, and
configuration of framework applications. Frameworks define processing
and programming models for a collaboration, as well as fulfill
requirements for interfacing to the computing model under which they
operate. The processing model is the mechanism used to execute and
apportion work. Mechanisms for this are threads, tasks, heavy-weight
processes along with interprocess communication. Programming model
elements, such as a well-defined logical design, scheduling, and
interactions with multiple languages that permit efficient and
maintainable algorithms, are also dictated by the processing
framework. The programming model also defines a well-thought out
physical code layout that minimizes coupling of logically independent
functionalities and libraries. This eases maintenance, extension and
refactoring, which is inevitable over the lifetime of HEP
experiments. The framework should provide hooks for doing monitoring
and logging services for performance and other purposes.

We have identified two kinds of required behavior that have directly
affected overall software system architecture, and therefore the
organization of the framework itself:

\begin{itemize}
\item
    Throughput maximizing: here it is most important to efficiently
    move data through all the available resources (memory, storage,
    and CPU), maximizing the number of events that are processed. The
    workload management systems used by experiments on the grid work
    towards this goal.
\item
    Latency minimizing (or reducing): online and interactive use
    cases where imposing constraints on how long it takes to calculate
    an answer for a particular datum is relevant and
    important. Dataflow and transaction processing systems work
    towards this goal.
\end{itemize}

Whether or not these differences necessitate fundamentally different
software systems remains to be seen, but both concepts need to be
accommodated and have relevant impact on the framework architecture
and design. Ways of accommodating both goals through the same system
architecture, or same software system with different configurations is
an area of research, and could enable us to meet some of the
challenges introduced in the previous section.

Current practice for throughput-maximizing system architectures have
constrained the scope of framework designs. Framework applications
have largely been viewed by the system as a batch job with complex
configuration, consuming resources according to rules dictated by the
computing model: one process using one core on one node, operating
independently with a fixed size memory space on a fixed set of files
(streamed or read directly). Only recently has CMS broken this
tradition starting in the beginning of Run 2, by utilizing all cores
on one virtual node in one process space using threading. ATLAS is
currently using a multi-process fork-and-copy-on-write solution to
remove the constraint of one core/process, and is now moving to the
multithreading approach too. Both experiments were
driven to solve this problem by the ever growing needs for more
memory per process brought on by the increasing complexity of LHC
events. Current practice manages system-wide (or facility-wide)
scaling by dividing up datasets, generating a framework application
configuration, and scheduling jobs on nodes/cores to utilize all
available resources. Given anticipated changes in hardware
(heterogeneity, connectivity, memory, storage) available at large
computing facilities, the interplay between workflow/workload
management systems and framework applications needs to be carefully
examined. It may be advantageous to permit framework applications (or
systems) to span resources, permitting them to be first-class
participatents in the business of scaling within a facility. O2 provides
a successful proof-of-principle of this approach.

\section{Roadmap}
\label{sec:roadmap}
Forward-looking work is underway as part of projects funded through government agencies, laboratories, and collaborations.  We want to be sure that relevant ideas and accomplishment are known, and that the groups doing this work have a place to report to and receive feedback for everyone’s benefit.
To organize the community, one needs to establish regular working group meetings, on a bi-monthly basis as we did with the concurrency forum. Face to face workshops after at least the 1st and the 3rd year can be co-hosted with events like CHEP and/or the WLCG workshops. A future planning workshop for transforming the results of the R\&D activities into a full development and deployment project plan should happen at the 5-year timescale.

\subsection{One-year goals}
\label{sec:one-year-goals}

Our assumption is that a set of one-year goals will be completed by
the end of 2018. We want work completed here to produce results that
will be useful in refining and moving into satisfying the three-year
goals.  Below is a list of all the areas that we believe need work and
ought to be considered when establishing projects that will help meet
all or some subset of these goals. Since this is R\&D, overall goals
are papers, workshop reports, analysis results, and software
architectures. A major purpose being to set direction for further
R\&D, or defining a path towards a new production product development,
or integration of results into existing software components.

\paragraph{Concept refinement} Jointly identify key abstractions that
made frameworks good for HEP in detail beyond what can be described in
this paper. Identify and describe where individual frameworks have
similarly or uniquely implemented these concepts. It is important to
describe how these choices are connected to the concrete use-cases. A
publishable paper should come of this that will serve as an agreed-upon
guide for where we can hope to go.

\paragraph{Technology investigations} There are four key areas that
ought to be explored to help determine future direction with regards
to software technology. The areas are: (1) task-based programming tools,
(2) inter-process and inter-node communication tools, (3) parallel number
crunching libraries, and (4) framework workflow management.

\paragraph{Functional programming} Conduct a study describing where we
currently are with functional programming. The study should address
the following questions. 1. What are the perceived benefits and
drawbacks?  2. Where can it be useful? 3. What does it mean in terms
of framework changes? 4. Where is language support headed for this
kind of programming, and recommendations for where to go next.

\paragraph{Support multiple concurrent scheduling tools, strategies,
  time scales, and granularities} Describe how an event-parallel
execution framework handles event-asynchronous I/O and processing
(e.g. alignment, offloading). Describe how underlying concurrency
libraries and tools can be shared. Describe how data blocks having
different granularities (e.g.  groups of particles, events) are
processed concurrently and synchronized. DNN (including simulation
GANs) offloading to GPU could be an early example to work on. A
parallelized Geant particle transport mechanism would be another one.

\paragraph{Domain-specific language} We believe it is useful to
describe the language that HEP uses to define and describe how to
simulate and reconstruct physics events. By language, we mean the
terms used to communicate and express how tasks are described and
carried out within a framework. This includes not only expressing data
dependencies, but also resource preferences and constraints, such as
GPUs. The goal here is to provide enough information for a group to
take on development of domain-specific libraries components and tools
that will increase the efficiency of carrying out physics. A good example
is how ML toolkits have evolved over the past few years. The
abstractions that have been developed have greatly increased productivity
and growth in the ML space such as the abstractions in Tensor Flow that
allow a coding of the matrix algebra that then gets remapped internally to
match the shape of the data being operated on.  The user only has take care
of getting the domain science functions correct.

\paragraph{Concluding workshop} Soon after the end of the first year,
our goal is to have a workshop. This workshop will be used for
reporting on results, reviewing alignment with written three-year
goals, and agreeing upon and developing next steps forward.

\subsection{Three-year goals}
\label{sec:three-year-goals}

Our assumption is that a set of three-year goals will be completed by
the end of 2020. Since LHC run three is targeted to start at this
time.  At least some of the focus ought to be on completing goals that
demonstrate advances over what will be in production during this
period.  An ongoing goal during this period ought to be incorporating
advancements in practices and tools into codes that will be available
during Run 3.

\paragraph{Common feature definition} Produce reports on how
frameworks ought to evolve to incorporate features of functional
programming, scheduling across heterogeneous resources, polyglot
programming, and addressing both necessary data model changes and I/O
handling.

\paragraph{Technology upgrades} Useful technologies identified during
the first phase ought to be further demonstrated and incorporated into
existing tools wherever possible.

\paragraph{Keeping up with evolving facilities} Memory, storage, and
network changes will be introduced during this timeframe. We want
studies that show how we might benefit from these changes. A goal is
to provide recommendations for how to react to ongoing facility
upgrades in the three areas listed above.

\paragraph{Common library integration} Plans for breaking out common
tools should be made during this period, along with demonstrating how
current frameworks might evolve to share more components.

\paragraph{Geant-framework cohesion} Make the next steps in bringing
underlying technology, toolkits, and designs together so Geant fits
well within the framework. A demonstration should be available in at
least one of the frameworks used in Run 3 of the LHC.

\paragraph{Blending of Workflow Management System functions}
Multi-node scheduling and heterogeneous resource management will start
to become more relevant during this period. Frameworks ought to share
responsibility with larger orchestration systems, and better exchange
information to increase flexibility in scheduling on diverse platforms
and architectures.

\paragraph{Facilities embedding} We have a consensus that the main
programming language we will be using in the near future is C++.  Much
of our community stresses use of modern language features. However,
some supercomputing facilities require the compilation of software
with dedicated compiler setups, thus making use of new language
features difficult. By the time of the milestone, we should have
members of the HEP community embedded in the decision-making process
for the provisioning of these machines. Since other programming
languages are also used, the same arguments can be made for them.

\paragraph{Continued coupling with ongoing R\&D activities} Projects
funded outside this work will continue through this period, and our
one-year goals ought to continue here.

\paragraph{Progress workshops} Workshops should happen at the end of
each year, with goals similar to those listed in the previous section.

\subsection{Five-year goals}
\label{sec:five-year-goals}

Our assumption is that a set of five-year goals will be completed by
the end of 2022. By this time we ought to have in place plans for
taking the R\&D results and incorporating them into the framework
software that will be further developed and utilized for HL-LHC. We
anticipate to work on these items:

\paragraph{Facilities readiness} A proven strategy for the integration
of HEP software frameworks with supercomputer centres and cloud
providers.  This work will be done in cooperation with the facilities
WG.

\paragraph{Common tools and practices demonstrated and in place where
  possible} Based on the common tools and components identified during
the \emph{common library integration} milestone, work on production
quality framework libraries for use by multiple experiments.

\paragraph{Interaction with Workflow and Data Management Systems} By
this time well-defined interfaces between those systems and frameworks
should be prepared, including at least one proven reference
implementation.

\paragraph{Incorporation of results from ongoing R\&D activities}
There will be independent progress on parallelization strategies and
implementations. At the time scale of 5 years we anticipate at least
one major paradigm shift to take place, which can not be incorporated
by continuous adjustment alone.

\paragraph{Defining what happens next} Based on the results of the
mentioned activities and the results of the other HSF working groups,
re-evaluate and define a concrete strategy for common framework
implementations for the coming years.

\subsection{Working towards these goals}
\label{sec:towards-goals}

To organize the community and to work towards these goals, one needs
to establish regular working group meetings, on a bi-monthly basis as
we did with the concurrency forum. Face to face workshops after at
least the 1st and the 3rd year can be co-hosted with events like CHEP
and/or the WLCG workshops. A future planning workshop for transforming
the results of the R\&D activities into a full development and
deployment project plan should happen at the 5-year timescale.

\sloppy
\raggedright
\clearpage
\printbibliography[title={References},heading=bibintoc]

\end{document}